\documentclass[11pt]{article}
\pdfoutput=1
\usepackage[utf8]{inputenc}
\usepackage[T1]{fontenc}
\usepackage{moriond,amsmath,lmodern}

\hypersetup{citecolor=blue}

\newcommand{\Photo}{}

\newcommand{\ugen}{U(3_\text{gen.})}
\newcommand{\ferm}{{5_\text{ferm.}}}
\newcommand{\gf}{\ensuremath{S\ugen^\ferm}}

\begin{document}
\vspace*{4cm}
\title{\uppercase{Flavourful baryon and lepton number violation\\at the LHC}\,\footnote{Talk given at the \emph{48th Rencontres de Moriond}, La Thuile, March 2013, and based on a work carried out in collaboration with J.-M. Gérard, F. Maltoni and C. Smith.\cite{Durieux:2012gj}}
}

\author{Gauthier Durieux}
\address{Centre for Cosmology, Particle Physics and Phenomenology (CP3), 
Université catholique de Louvain, 
Chemin du Cyclotron 2, 
B-1348 Louvain-la-Neuve, 
Belgium
}

\maketitle
\abstracts{We observe that the flavour symmetries of the Standard Model gauge sector, broken as they are in the Standard Model Yukawa Lagrangian, naturally suppress baryon and lepton number violation at low energies and, simultaneously, make it accessible at the LHC through resonant processes involving at least six fermions from all three generations. We establish a model independent classification of such transitions and identify two classes that give rise to particularly clean LHC signatures, namely $\bar t\, \mu^+\, e^+$ and $\bar t\, \bar t\:+$~jets.
}

\vspace*{-2mm}
\section{Introduction}
\vspace*{-2mm}
The global baryon ($B$) and lepton number ($L$) symmetries of the Standard Model (SM) Lagrangian cannot be absolute. Theoretical support to this statement come from multiple sources. Quantum effects within the SM itself notably allow for non-perturbative transitions that violate both $B$ and $L$ (only $B-L$ is anomaly-free). Those effects are nonetheless negligible at low temperatures and energies. Beyond the SM, it is also very difficult to naturally reproduce an accidental conservation that is not justified by any fundamental principle and disappears as soon as the matter content or the gauge group of the SM is altered.

However, more than half a century of experimental tests at low energies, most notably for nucleon decay,\footnote{Reines, Cowan and Goldhaber\,\cite{Reines:1954pg} were probably the firsts to carry out a dedicated search in 1954.} have been so far unsuccessful in discovering any baryon and lepton number violation (BLV). Still, indirect evidences are provided by the observed baryon--anti-baryon asymmetry of the Universe (that would require $B$ violation to have occurred at some moment of its history) and by the tiny neutrino masses that are commonly accounted for by making neutrinos Majorana particles (therefore introducing explicit $L$ violation in the SM Lagrangian).

At the high-energy frontier, the LHC now offers us the opportunity of observing a so far elusive violation. The minimal effective field theory for BLV,\cite{Dong:2011rh} without flavour requirement, has already been searched for at the LHC.\cite{CMS:2012pxa} If assumed though, the SM flavour symmetries actually constrain BLV to involve at least six fermions belonging preferentially to all three generations. Significant suppression then affect other flavour structures like the ones restricted to light generations only that are of primary relevance in low-energy processes. The non-observation of low-energy BLV may therefore not be as unexpected as it seems at first sight and BLV may not be intrinsically small. A natural compliance with low-energy constraints can actually be obtained with a scale for BLV in the TeV range so that resonant processes violating $B$ and $L$ with signature such as $\bar t\, \mu^+\, e^+$ or $\bar t\,\bar t\,+$~jets could be seen at the LHC.

\vspace*{-2mm}
\section{Standard Model flavour symmetries}
\vspace*{-2mm}
As pointed out by Chivukula and Georgi,\cite{Chivukula:1987py} the SM gauge sector has a large global symmetry:
\begin{equation*}
\begin{aligned}
\\[-8mm]
\ugen^\ferm
& = \ugen_q \times \ugen_u \times \ugen_d \times \ugen_l \times \ugen_e
\\
& = S\ugen^\ferm \times U(1)_Y \times U(1)_B \times U(1)_L \times U(1)_d \times U(1)_e
\end{aligned}
\end{equation*}
as each of the five fermionic fields ($q,\,l,$ the quark and lepton doublets, $u,\,d,\,e,$ the quark and lepton singlets) can be independently and unitarily rotated in generation space without altering the gauge Lagrangian.
The non-Abelian $SU(3)$'s and Abelian $U(1)$'s of this large global symmetry are to be treated separately since global $U(1)$'s are often found to be anomalous. Taking combinations of the original $U(1)_{q,u,d,l,e}$, we can actually identify one Abelian factor for the hypercharge $Y$ (which will be promoted to a gauge symmetry) as well as two $U(1)$'s for the baryon and lepton number which are known to be anomalous.

The non-Abelian part \gf\ or \emph{flavour symmetry} group is explicitly broken in the Yukawa sector of the SM. Though, this explicit breaking is very specific: charged current mixing different generations are usually small since the Cabibbo–Kobayashi–Maskawa (CKM) matrix is very hierarchical and flavour changing neutral currents (FCNC) are almost perfectly vanishing because of the Glashow-Iliopoulos-Maiani (GIM) mechanism. Any deviation from this SM picture of flavour transitions is moreover very well constrained experimentally.
In the following, we are going to examine how a SM-like breaking of \gf\ constrains BLV. Such a breaking can for instance be explicitly implemented using the Minimal Flavour Violation (MFV) hypothesis but we would like to draw conclusions as general as possible and will not place ourselves in this framework. Note the intriguing fact that $U(1)_{B,L}$ and \gf\ seem to have a common origin in $\ugen^\ferm$. This may suggest to consider $B$- and $L$-violating processes as flavour transitions of a special kind on which constraints arising from the already precisely measured flavour structures naturally derive.

\vspace*{-2mm}
\section{Three-generation baryon and lepton number violation}
\vspace*{-2mm}
Before breaking it in a SM-like fashion, let us thus impose the strict conservation of \gf\ together with BLV. The two $SU(3)$-invariant tensors we can make use of are $\delta^a_b$ and $\epsilon_{abc}$. However, structures based on Kronecker delta as $\delta^a_b\:\bar \psi_a\psi^b$ do not lead to any BLV. The elementary building block of a \gf-invariant BLV is therefore of the form $\epsilon_{abc}\psi^a\psi^b\psi^c$ (or $\epsilon^{abc}\bar\psi_a\bar\psi_b\bar\psi_c$). Moreover, Lorentz invariance, that requires the number of fermions involved in any process to be even, asks for at least two of these completely anti-symmetrical flavour structures to be combined together. With exact \gf\ invariance, we therefore note that a minimal content of at least six fermions is required and that all three generations are involved simultaneously in $B$- and $L$-violating processes. In such a setup, the presence of heavy generations kinematically forbids any nucleon decay.

Though, imposing the exact conservation of \gf\ is not a realistic assumption since this symmetry is already broken by SM Yukawa interactions (as unambiguously manifest from fermion masses). Because any deviation from this specific breaking scheme is very well constrained experimentally, let us only introduce SM-like breaking terms. These Yukawas mix left- and right-handed quarks or leptons but not quarks and leptons together. Moreover, changes of generations are usually costly since the CKM off-diagonal matrix elements are quite small and FCNC incredibly suppressed. Therefore, a SM-like breaking of the \gf\ symmetry still requires BLV to involve at least two combinations of three (anti-)\linebreak[2]quarks or three (anti-)\linebreak[2]leptons. The simultaneous presence of all three generations is however not any longer absolutely required. Still, where kinematically allowed, the unsuppressed completely antisymmetrical flavour structures will dominate over the other ones. 

With this minimal number of fields (namely six-leptons, three leptons and three quarks, or six quarks) we note that only four selections rules are allowed: $(\Delta B,\Delta L)=(0,\pm 6)$, $(\pm 1,\pm 3)$, $(\pm 1,\mp 3),\, \text{ and } (\pm 2,0)$.

\vspace*{-2mm}
\section{Implications at low energies}
\vspace*{-2mm}
At energies much lower than its characteristic scale, a BLV compatible with SM flavour symmetries can conveniently be described by an effective field theory. Since the minimal number of fields involved is of six fermions, dimension-six operators\,%
\cite{Weinberg:1979sa}
 are absent and the ones present are at least suppressed by the BLV scale to the fifth power. Moreover, the presence of all three generations is favoured and processes involving light generations only are expected to suffer from significant flavour suppressions. The latter can be explicitly computed if the MFV framework is used to implement a SM-like breaking of \gf.\cite{Smith:2011rp} The combination of these dimensional and flavour suppressions cause the limits set on nucleon decay or neutron anti-neutron oscillation to translate into a bound on the BLV scale in the TeV range.\cite{Smith:2011rp}

\vspace*{-2mm}
\section{Implications at high energies}
\vspace*{-2mm}
Resonant processes could therefore be seen at high-energy colliders. Describing them quantitatively in a model independent way is however not an easy task as non-local processes cannot be modelled in an effective field theory framework (explicit simplified models were presented elsewhere\,\cite{Durieux:2012gj}). Though, the reasoning leading to the four selection rules for $(\Delta B,\Delta L)$ remains valid since it was based on the use global of symmetries only.

In the resonant (non-local) regime, the adjunction of other $\Delta B=0=\Delta L$ combinations of fields in addition to the minimal six fermions cannot any longer be dismissed by calling for the fact that they would lead to processes that are further suppressed. In the effective-field-theory (local) regime only, do non-minimal operators suffer from more and more severe dimensional suppressions. The fermionic core that drives the BLV can be isolated and used to establish a classification of allowed processes but, for collider phenomenology, there is in principle no reason to restrict ourselves to the minimal field content. Still, in each class, the transitions involving a number of fields far above the minimal one would often be more difficult to observe experimentally. We will therefore address the simplest cases only and introduce slight deviations from the minimal field content where they lead to phenomenologically interesting signatures. The introduction of extra gluon fields or the interchange of a fermion with a $W$ and its $SU(2)_L$ partner are for instance considered. Finally, we should bear in mind that (as in the effective-field-theory regime) BLV may possibly occur only in processes involving a field content that we qualified as non-minimal here.

We have already used flavour and Lorentz symmetries to point at the minimal content of six fermions. We can further impose the overall conservation of electric charge (colour conservation does not give any extra constraint) to identify fermionic cores upon which an equal number of distinct classes of $B$- and $L$-violating processes derives. Those fermionic cores are listed in Table~\ref{tab:cores} together with explicit examples of the three-generation flavour structures that are relevant for $B$- and $L$-violating transitions at colliders.

\begin{table}[htb]
\newcommand{\ms}{\quad}
\newcommand{\p}{$+$} 
\newcommand{\m}{$-$} 
\newlength{\intercol}
\settowidth{\intercol}{\qquad}
\newcommand{\ic}{\hspace{\intercol}}
\begin{tabular*}{\textwidth}[c]{%
|c@{\;}c @{\ic\extracolsep{\fill}}
	r@{\,\,}l @{\ic}
		r@{\,\;\,}l @{\ic}
			l @{\ic}
				c|
					}\hline
$\Delta B$&$\Delta L$ 
	& \multicolumn{2}{c}{\hspace{-\intercol}Fermionic cores} 
		& \multicolumn{2}{c}{\hspace{-\intercol}Examples}
			& Promising LHC processes 
				& $A_{e\mu}$
					\\\hline
$0$&$\pm6$ 
	& NNN&NNN 
		& $\nu_{e}\,\nu_{\mu}\,\nu_{\tau}$&$\nu_{e}\,\nu_{\mu}\,\nu_{\tau}$
			& $u\,\bar{u}\to e^{-}\mu^{-}\nu_{\tau}\nu_{e}\nu_{\mu}\nu_{\tau}\ms W^{+}W^{+}\;$ 
				& $0$
				\\\hline
$\pm1$&$\pm3$ 
	& UUU&EEN
		& $t\,c\,u$&$e^{-}\,\mu^{-}\,\nu_{\tau}$
			& $u\,c\to\bar{t}\ms e^{+}\,\mu^{+}\,\bar{\nu}_{\tau}$ 
				& \p
				\\
&	&\omit&	&\omit&	& $g\,g\to\bar{t}\,\bar{c}\,\bar{u}\ms e^{+}\,\mu^{+}\,\bar{\nu}_{\tau}$
				& $0$
				\\
&	& UUD&ENN
		& $t\,c\,d$&$e^{-}\,\nu_{\mu}\,\nu_{\tau}$
			& $d\,c\to\bar{t}\ms e^{+}\,\mu^{+}\,\bar \nu_{\tau}\ms W^{-}$
				& \p
				\\
&	& UDD&NNN 
		& $t\,s\,d$&$\nu_{e}\,\nu_{\mu}\,\nu_{\tau}$
			& $d\,s\to\bar{t}\ms e^{+}\,\mu^{+}\,\bar \nu_{\tau}\ms W^{-}W^{-}$
				& \p
				\\\hline
$\pm1$&$\mp3$
	& UDD&\=N\=N\=N
		& $t\,s\,d$&$\bar\nu_{e}\,\bar\nu_{\mu}\,\bar\nu_\tau$
			& $d\,s\to\bar t \ms e^-\,\mu^-\,\nu_\tau\ms W^{+}W^{+}$
				& \m
				\\
&	& DDD&\=E\=N\=N
		& $b\,s\,d$&$e^{+}\,\bar\nu_\mu\,\bar\nu_\tau$
			& $d\,s\to\bar t \ms e^-\,\mu^-\,\nu_{\tau}\ms W^{+}W^{+}$
				& \m
				\\\hline
$\pm2$&$0$
	& UDD&UDD
		& $t\,s\,d$&$t\,s\,d$
			& $d\,d\to\bar{t}\,\bar{t}\ms \bar{s}\,\bar{s}$
				& \m
				\\
&	&\omit&	&\omit&	& $g\,g\to\bar{t}\,\bar{t}\ms \bar{s}\,\bar{s}\ms \bar{d}\,\bar{d}$
				& $0$
				\\
&	& &
		& $t\,c\,d$&$b\,s\,d$
			& $d\,d\rightarrow\bar{t}\,\bar{t}\ms\bar{c}\ms \bar{s}\ms W^+$
				& \m
				\\\hline
\end{tabular*}
\vspace*{-2mm}
\caption{Baryon and lepton number violation compatible with Standard Model flavour symmetries. Capital $N,\,E,\,D,\,U$ stand for flavour-generic neutrinos, charged leptons, down- and up-type quarks. Conjugate field contents with anti-fermions instead of fermions and \emph{vice versa} are understood.}
\vspace*{-6mm}
\label{tab:cores}
\end{table}

\vspace*{-2mm}
\section{LHC phenomenology}
\vspace*{-2mm}
At the LHC, the dominant three-generation signatures would be accessible through resonant processes as all fermions, even the heavy third-generation ones, are produced copiously.

One first comment that we can make with respect to the content of Table~\ref{tab:cores} is that the simplest examples of $B$- and $L$-violating processes compatible with SM flavour symmetries all involve \emph{same-sign} fermions \emph{i.e.} either quarks (leptons) or anti-quarks (anti-leptons) but not both simultaneously. As a consequence, BLV at the LHC would feature same-sign final states. To fully exploit this characteristic signature, the charges of final state particles need to be identified. This is most easily done for charged muons and electrons as well as for top quarks decaying semi-leptonically. The examples of fermionic cores $t\,c\,u\;\; e^-\mu^- \nu_\tau$ with $(\Delta B,\Delta L)=(\pm 1,\pm 3)$ and $t\,s\,d\;\; t\,s\,d$ with $(\Delta B, \Delta L)=(\pm 2, 0)$ that feature the largest number charged leptons and tops are thus of particular interest.
Without relying on a specific model, we can identify promising processes built upon their field content (see Table~\ref{tab:cores}). Doing so, it may be advantageous to put gluons in the initial state instead of quarks as this may lead to parton-distribution-functions enhancements or to a different resonant structure associated to higher signal rates. Replacing a neutrino by a charged lepton and a $W$ is another possible step beyond minimality that can yield more easily observable final states.

One second important feature of the LHC signature is the di-lepton charge asymmetry. As a proton-proton collider, the LHC has an asymmetric initial state characterised by a net baryon number equal to two and a significant predominance of $u$'s and $d$'s over other quarks. Processes initiated by these valence quarks therefore occur much more often than their conjugates involving initial $\bar u$'s and $\bar d$'s. Transitions $u\,c\to \bar t\; e^+\,\mu^+\,\bar\nu_\tau$ and $d\,d\to \bar t\,\bar t\;\bar s\,\bar s$ for instance lead to much more positively charged leptons and anti-tops pairs than negatively charged leptons and tops pairs, respectively. Given that anti-tops decay semi-leptonically to negatively charged leptons, a charge asymmetry in same-sign di-lepton $B$- and $L$-violating production defined as
\begin{equation*}
A_{\ell\ell'}\equiv \frac{\sigma^\text{BLV}(\ell^+\ell'^+) -\sigma^\text{BLV}(\ell^-\ell'^-)}{\sigma^\text{BLV}(\ell^+\ell'^+) +\sigma^\text{BLV}(\ell^-\ell'^-)}
\end{equation*}
would be positive in the first $(\Delta B,\Delta L)=(\pm 1,\pm 3)$ case and negative in the second $(\Delta B, \Delta L)=(\pm 2, 0)$ one (see Table~\ref{tab:cores}). This observable therefore provides an interesting discrimination power between different scenarios. Remarkably, the SM as well as almost all its new physics extensions but this baryon number violating one feature positive asymmetries only.\cite{Durieux:2012gj}

\vspace*{-2mm}
\section{Conclusions}
\vspace*{-3mm}
We have analysed, in a model independent way, baryon and lepton number violation compatible with Standard Model flavour symmetries. Its characteristic scale is allowed by low-energy constraints to lie no higher than the TeV range. Resonant transitions involving all three generations could therefore be observable at the LHC.

\vspace*{-2mm}
\section*{Acknowledgements}
\vspace*{-3mm}
This work has been carried out in collaboration with J.-M. Gérard, F. Maltoni and C. Smith who are warmly thanked here. The author is a Research Fellow of the F.R.S.-FNRS, Belgium.

\vspace*{-3mm}
\section*{References}
\vspace*{-3mm}
\begin{flushleft}
\bibliographystyle{apsrev4-1-moriond}
\bibliography{moriond.bib}
\end{flushleft}
\end{document}